\documentclass[twocolumn,nofootinbib]{revtex4}%

\usepackage{amssymb}
\usepackage{amsfonts}
\usepackage{amsmath}
\usepackage[latin1]{inputenc}
\usepackage{graphicx,color}
\usepackage{slashed}
\usepackage{young}
\usepackage[vcentermath]{youngtab}
\usepackage{tensor}

\makeatletter

\newcommand{\be}{\begin{equation}}
\newcommand{\ee}{\end{equation}}
\newcommand{\bea}{\begin{eqnarray}}
\newcommand{\eea}{\end{eqnarray}}

\newcommand{\del}{\partial}

\def\nn{\nonumber} 

\numberwithin{equation}{section}

   \def\m{\mu}
    \def\r{\rho}
\def\s{\sigma}

 \def\cB{{\cal B}} \def\cC{{\cal C}} 
 \def\cE{{\cal E}}  
 \def\cH{{\cal H}}  
   
\def\cM{{\cal M}}   
  \def\cR{{\cal R}} 
\def\cS{{\cal S}}  
 \def\cW{{\cal W}}

\begin{document}

\title{Manifest electric-magnetic duality in linearized conformal gravity}
\author{Hap\'e Fuhri Snethlage, Sergio H\"ortner}
\email{sergio.hortner@uva.nl}
\affiliation{\medskip
Institute for Theoretical Physics, University of Amsterdam,\bigskip
Science Park 904, 1090 GL Amsterdam, The Netherlands}

\begin{abstract}
We derive a manifestly duality-symmetric formulation of the action principle for conformal gravity linearized around Minkowski space-time. The analysis is performed in the Hamiltonian formulation, the fourth-order character of the equations of motion requiring the formal treatment of the three-dimensional metric perturbation and the extrinsic curvature as independent dynamical variables. The constraints are solved in terms of two symmetric potentials that are interpreted as a dual three-dimensional metric and a dual extrinsic curvature. The action principle can be written in terms of these four dynamical variables, duality acting as simultaneous rotations in the respective spaces spanned by the three-dimensional metrics and the extrinsic curvatures. A twisted self-duality formulation of the equations of motion is also provided.

\end{abstract}

\maketitle

\section{Introduction}
\setcounter{equation}{0}


Understanding dualities is a major challenge in modern theoretical physics. Despite their widespread presence in diverse branches --including field theory, (super)gravity, string theory and condensed matter-- a comprehensive theoretical framework that explains the origin of the phenomenon and describes its full implications is lacking. In the case of gravitational theories, dualities are intimately related to the emergence of hidden symmetries upon toroidal compactifications (supplemented by Hodge dualizations of Kaluza-Klein fields). For instance, it has long been recognized that the reduction to three dimensions of the four-dimensional Einstein-Hilbert action in the presence of a Killing vector, followed by the dualization of the Kaluza-Kelin vector to a scalar, exhibits a $SL(2,R)$ invariance acting on the scalar sector. The latter, conformed by a dilaton and axion, parametrizes a $SL(2,R)/SO(2)$ coset space. This $SL(2,R)$ invariance is commonly referred to in the literature as the Ehlers symmetry \cite{Ehlers}. In the presence of two commuting Killing vectors, the reduction to two dimensions can be achieved in two different ways: either by a direct compactification to two dimensions, or by a reduction to three dimensions followed by a dualization of the vector field to a scalar and a final reduction to two dimensions. Each of these routes yields a different $SL(2,R)/SO(2)$ sigma model in the scalar sector. The first one is associated to the $SL(2,R)$ Matzner-Misner group, which originates from coordinate transformations preserving the vector space spanned by the Killing vectors. In the second case the $SL(2,R)$ group is properly the Ehlers group. The intertwining of the Matzner-Misner and Ehlers groups generates an infinite-dimensional group realized non-locally, as described by Geroch \cite{Geroch}. 

\medskip

A similar situation occurs in supergravity \cite{Cremmer}: the (11-$d$) toroidal compactification of the eleven-dimensional theory yields maximally supersymmetric supergravity in dimension $d$ with a symmetry structure hidden within the non-gravitational degrees of freedom in the reduced bosonic sector, namely $p$-forms and scalars. After Hodge dualization of the $p$-forms to their lowest possible rank, they combine in an irreducible representation of a non-compact group $G$ acting globally, whereas the scalar sector is described by the non-linear sigma model $G/H$, with $H$ the maximal compact subgroup of $G$. In even dimensions, the global symmetry $G$ is realized as an electric-magnetic duality transformation interchanging equations of motion and Bianchi identities. Reduction to five, four and three dimensions yields as $G$ the exceptional Lie groups $E_{6(6)}$, $E_{7(7)}$ and $E_{8(8)}$, respectively. 

\medskip

The study of the rich algebraic structure that underlies the emergence of hidden symmetries in compactifications of (super)gravity has led to conjecture the existence of an infinite-dimensional Kac-Moody algebra acting as a fundamental symmetry of the uncompactified theory \cite{Julia}-\cite{Lambert}, encompassing the duality symmetries that appear upon dimensional reduction. A key property of these algebras is that they involve all the bosonic fields in the theory and their Hodge duals, including the graviton and its dual field. The associated symmetry transformation for a given tensor field relates it to all the rest of the fields, regardless their tensor structure, and in general has a highly non-trivial form. In four dimensions, the graviton and its dual field are each described by a symmetric rank-two tensor field, and a duality symmetry relating them is expected to emerge, inherited from the underlying infinite-dimensional algebraic structure. This has motivated the search of duality-symmetric action principles involving gravity \cite{HT}-\cite{me}, along the lines of the work \cite{deser1} establishing a manifestly duality symmetric formulation of Maxwell action.

\medskip

It seems natural to wonder about the possibility of deriving duality-symmetric action principles for theories of gravity involving higher derivatives. Among those, conformal gravity occupies a position of particular interest in the literature. Being constructed out of the square of the Weyl tensor, the action principle is invariant under conformal rescalings of the metric. As opposed to Einstein gravity, it is power-counting renormalizable \cite{stelle},\cite{adler}, albeit it presents a Ostrogradski linear instability in the Hamiltonian --due to the fourth-order character of the equations of motion and the non-degeneracy of the Lagrangian \footnote{Recall that a higher order Lagrangian is non-degenerate whenever the highest time derivative term can be expressed in terms of the canonical variables.}--, which is typically assumed to translate into the presence of {\itshape{ghosts}} --negative-norm states-- upon quantization. It is well known that solutions of Einstein gravity form a subset of solutions of conformal gravity, a fact that has recently been exploited in \cite{maldacena} to show the equivalence at the classical level of Einstein gravity with a cosmological constant and conformal gravity with suitable boundary conditions that eliminate ghosts. Another interesting aspect of conformal gravity is that it admits supersymmetric extensions for ${\cal{N}}\leq 4$, the maximally supersymmetric theory admitting different variants (see \cite{susy1} for a review and \cite{dewitt} for recent progress). Other theoretical advances involving conformal gravity include its emergence from twistor string theory \cite{witten1} and its appearance as a counterterm in the AdS/CFT correspondence \cite{adscft}. 

\medskip

A generalization of electric-magnetic duality in conformal gravity was studied in the early work \cite{deser}, where the Euclidean action with a gauge-fixed metric was expressed in terms of quadratic forms involving the electric and magnetic components of the Weyl tensor, exhibiting a discrete duality symmetry upon the interchange of these components. This result can be regarded as the analog of the duality symmetry of Euclidean Maxwell action under the exchange of electric and magnetic fields. Unlike \cite{deser1}, duality is discussed in terms of the electric and magnetic components of the curvature, and not at the level of the dynamical degrees of freedom of the theory. 
\medskip

In the present article we focus on linearized conformal gravity with Lorentzian signature and show that the action principle admits a manifestly duality invariant form in terms of the dynamical variables. The derivation requires working in the Hamiltonian formalism, the identification of the constraints --both algebraic and differential-- and the resolution of the differential ones in terms of potentials, that we will eventually interpret as a dual metric and a dual extrinsic curvature. The structure of the duality-symmetric action principle is new, different from duality-invariant Maxwell theory and linearized gravity: duality acts rotating simultaneously the three-dimensional metrics $(h_{ij}, \tilde{h}_{ij})$ and the extrinsic curvatures $(K_{ij},\tilde{K}_{ij})$. 

\medskip

The rest of the article is organized as follows. In Section II we review general features of conformal gravity and remark that, in the linearized regime, the Hodge dual of the linearized Weyl tensor obeys an identity of the same functional form as the equation of motion satisfied by the Weyl tensor itself, in complete analogy with the symmetric character of vacuum Maxwell equations with respect to the exchange of the field strength and its Hodge dual. Motivated by this observation, in Section III we establish a twisted self-duality form of the linearized equations of motion of conformal gravity. Section IV deals with the generalities of the Hamiltonian formulation, including the identification of the dynamical variables and the constraints of the theory. To deal with the fact that the Lagrangian contains second order time derivatives of the metric perturbation, we will formally promote the linearized extrinsic curvature to an independent dynamical variable. Section V is dedicated to the resolution of the differential constraints in terms of two potentials. These are interpreted as a dual three-dimensional metric and a dual extrinsic curvature. In Section VI we present a manifestly duality invariant form of the action principle, where the two metrics and extrinsic curvatures appear on equal footing. Finally we draw our conclusions in Section VII and set out proposals for future work.

\section{Conformal gravity}
\setcounter{equation}{0}

The action principle of conformal gravity is given by

\begin{equation}
S[g_{\mu\nu}]=-\frac{1}{4}\int\;d^{4}x\sqrt{-g}\cW_{\mu\nu\r\s}\cW^{\mu\nu\r\s}\label{actionf},
\end{equation}
with $g_{\mu\nu}$ the metric tensor defined on a manifold $\cM$ and $\cW^{\mu}_{\ \nu\rho\sigma}$ the Weyl tensor

\begin{equation}
\tensor{\cW}{^\mu_\nu_\r_\s}\equiv \tensor{\cR}{^\mu_\nu_\r_\s}-2(g^{\mu}_{\ [\r}\cS_{\s]\nu}-g_{\nu[\r}\cS_{\s]}^{\ \mu}).
\end{equation}
Here ${\cal{R}}^{\mu}_{\ \nu\rho\sigma}$ and ${\cal{S}}_{\mu\nu}$ are the Riemann and Schouten tensors, respectively. The latter is defined as

\begin{equation}
\cS_{\mu\nu} \equiv \frac{1}{2}(\cR_{\mu\nu}-\frac{1}{6}g_{\mu\nu}\cR).
\end{equation}
We adopt the convention that indices within brackets are antisymmetrized, with an overall factor of $1/n!$ for the antisymmetrization of $n$ indices.

\medskip

The Weyl tensor $\cW^{\mu}_{\ \nu\r\s}$ is invariant under diffeomorphisms 

\begin{equation}
\delta g_{\mu\nu} = \nabla_{\mu}\xi_{\nu} + \nabla_{\nu}\xi_{\mu},
\end{equation}
and local conformal rescalings of the metric

\begin{equation}
g_{\mu\nu}\rightarrow g'_{\mu\nu}=\Omega^2(x) g_{\mu\nu}.
\end{equation}
These transformations also determine the symmetries of the action principle (\ref{actionf}). 

\medskip

The Weyl tensor satisfies the same tensorial symmetry properties as the Riemann tensor,

\begin{align}
\cW_{\mu\nu\r\s}=-\cW_{\nu\mu\r\s}=-\cW_{\mu\nu\s\r}=\cW_{\r\s\mu\nu},\label{symm}
\end{align}
as well as the identities

\begin{eqnarray}
&\cW_{[\mu\nu\s]\r}=0,\label{id}
\end{eqnarray}

\begin{eqnarray}
&\cW^{\mu}_{\ \nu\mu\s}=0,\label{traceless}
\end{eqnarray}
and

\begin{equation}
\nabla_{\mu}\tensor{\cW}{^\mu_\nu_\r_\s}=-{\cal{C}}_{\nu\r\s},\label{bianchi}
\end{equation}
where we have introduced the Cotton tensor
\begin{align}
&\cC_{\nu\r\s}\equiv 2\nabla_{[\r}\cS_{\s]\nu}.
\end{align}
Equation (\ref{bianchi}) is a consequence of the Bianchi identity for the Riemann tensor.

\medskip

The fourth-order equation of motion derived from the conformal gravity action principle (\ref{actionf}) reads


\begin{equation}
(2\nabla_{\r}\nabla^{\s} + \cR_{\r}^{\ \s})\cW^{\r}_{\ \mu\s\nu}=0.\label{eomf}
\end{equation}
This is usually referred to as the Bach equation, the left-hand side of (\ref{eomf}) being dubbed the Bach tensor. Clearly, conformally flat metrics constitute a particular subset of solutions to the equations of motion (\ref{eomf}). Einstein metrics constitute another subset of particular solutions.


\medskip

\begin{center}
{\bf{Remarks on the linearized regime}}
\end{center}

In the linearized regime

\begin{equation}
g_{\mu\nu} = \eta_{\mu\nu} + h_{\mu\nu}
\end{equation}
the Weyl tensor takes the form

\begin{align}
&W^{\mu}_{\ \nu\rho\sigma}[h] = \tensor{R}{^\mu_\nu_\r_\s}[h]-2(\delta^{\mu}_{\ [\r}S_{\s]\nu}[h]-\delta_{\nu[\r}S_{\s]}^{\ \mu}[h]),\label{weyltensor}
\end{align}
where $R_{\mu\nu\r\s}$ is the linearized Riemann tensor 

\begin{align}
R_{\mu\nu\r\s}=-\frac{1}{2}[\partial_{\mu}\partial_{\r}h_{\nu\s}+\partial_{\nu}\partial_{\s}h_{\mu\r}-\partial_{\mu}\partial_{\s}h_{\nu\r}-\partial_{\nu}\partial_{\r}h_{\mu\s}]\label{riemann}
\end{align}
and $S_{\mu\nu}$ the linearized Schouten tensor. The action principle and equations of motion reduce to

\begin{align}
&S[h_{\alpha\beta}]=-\frac{1}{4}\int\;d^{4}xW_{\mu\nu\r\s}[h]W^{\mu\nu\r\s}[h]
\end{align}

and

\begin{equation}
\del_{\mu}\del_{\nu}W^{\mu\r\nu\s}[h]=0.
\label{eoml}
\end{equation}
The linearized Weyl tensor still obeys the symmetry properties (\ref{symm}), and the identity (\ref{bianchi}) takes the linearized form

\begin{equation}
\del_{\mu}W^{\mu\nu\r\s}[h]=-C^{\nu\r\s}[h].
\label{weyl_cotton}
\end{equation}
This allows for a rewriting of the linearized Bach equation in terms of the Cotton tensor:

\begin{equation}
\del_{\r}C^{\nu\r\s}[h]=0.
\end{equation}

Let us now introduce the Hodge dual of the linearized Weyl tensor:

\begin{equation}
^{*}W_{\mu\nu\rho\sigma}[h]\equiv\frac{1}{2}\tensor{\epsilon}{_\mu_\nu^\alpha^\beta}\tensor{W}{_\alpha_\beta_\r_\s}[h].
\end{equation}
By construction it possesses the same symmetries as $W_{\mu\nu\r\s}$, namely
\begin{equation}
^{*}W_{\mu\nu\r\s}=^{*}W_{\r\s\mu\nu}=-^{*}W_{\nu\mu\r\s}=-^{*}W_{\mu\nu\s\r}.
\end{equation}
It also satisfies the cyclic identity

\begin{equation}
^{*}W_{[\mu\nu\r]\s}=0
\end{equation}
and is traceless

\begin{equation}
^{*}W^{\mu}_{\ \nu\mu\r}=0.
\end{equation}
At this point, it is crucial to observe that $^{*}W_{\mu\nu\rho\sigma}$ satisfies the following identity:

\begin{equation}
\del_{\mu}\del_{\nu}\;^{*}W^{\mu\r\nu\s}[h]=0.\label{id_dual_w}
\end{equation}
This is directly related to the identity 

\begin{equation}
C_{\s[\nu\r,\mu]}[h]=0
\end{equation}
satisfied by the linearized Cotton tensor, for

\begin{eqnarray}
&\del_{\mu}\del_{\nu}\;^{*}W^{\mu\r\nu\s}[h]=\frac{1}{2}\del_{\mu}\del_{\nu}\epsilon^{\mu\r\alpha\beta}W_{\alpha\beta}^{\ \ \nu\s}[h]\nn\\
&=-\frac{1}{2}\del_{\mu}\epsilon^{\mu\r\alpha\beta}C^{\s}_{\ \alpha\beta}[h]=0.
\end{eqnarray}
It is now clear that the set of equations conformed by the linearized Bach equation (\ref{eoml}) and the identity (\ref{id_dual_w})

\begin{align}
\partial_{\mu}\partial_{\rho}W^{\mu\nu\rho\sigma}[h]&=0\nn\\
\partial_{\mu}\partial_{\rho}\,^{*}W^{\mu\nu\rho\sigma}[h]&=0\label{WWW}
\end{align}
may be regarded as the analog of Maxwell equations in vacuum

\begin{align}
\partial_{\mu}F^{\mu\nu}[A]&=0\nn\\
\partial_{\mu}\, ^{*}F^{\mu\nu}[A]&=0.\label{electrom}
\end{align}
The set of equations (\ref{WWW}) is symmetric under the replacement of the Weyl tensor and its Hodge dual.

\section{Twisted self-duality form of the equations of motion}
\setcounter{equation}{0}

Given the formal resemblance between equations (\ref{electrom}) and (\ref{WWW}), it seems natural to wonder about the existence of an underlying electric-magnetic duality structure in linearized conformal gravity. In this section we show that the set of equations (\ref{WWW}) can be cast in a covariant twisted self-duality form, and that the non-covariant subset defined by selecting the purely spatial components of the latter contains all the information of the full covariant set --which parallels the situation in electromagnetism \cite{abh} and linearized Einstein gravity \cite{mebh}. 

\medskip

In order to understand the logic underlying twisted self-duality, it is useful to briefly recall the situation in Maxwell theory. Consider the vacuum equations (\ref{electrom}), where we tacitly assume $F_{\mu\nu}=\partial_{\mu}A_{\nu}-\partial_{\nu}A_{\mu}$. Although (\ref{electrom}) are symmetric under the exchange of $F_{\mu\nu}[A]$ and $^{*}F_{\mu\nu}[A]$, these quantities do not appear exactly on an equal footing: the equation for $^{*}F_{\mu\nu}$ is an identity. In other words, $^{*}F_{\mu\nu}$ has been implicitly solved in terms of the potential $A^{\mu}$. Indeed, upon use of Poincar\'e lemma, one finds $^{*}F_{\mu\nu}=\epsilon_{\mu\nu\alpha\beta}\partial^{\alpha}A^{\beta}$ for some vector potential $A^{\mu}$, and the definition of the Hodge dual yields $F_{\mu\nu}=\partial_{\mu}A_{\nu}-\partial_{\nu}A_{\mu}$, as expected. We seek instead a formulation where $F_{\mu\nu}$ and $^{*}F_{\mu\nu}$ appear on equal footing, with no implicit prioritization of any of them. This is achieved \cite{abh} by considering the field strength and its Hodge dual as independent variables, solving simultaneously for both in terms of potentials $F_{\mu\nu}[A]=\partial_{\mu}A_{\nu}-\partial_{\nu}A_{\mu}$ and $^{*}F_{\mu\nu}\equiv H_{\mu\nu}[B]=\partial_{\mu}B_{\nu}-\partial_{\nu}B_{\mu}$, and finally imposing a first-order, twisted self-duality condition that takes into account that $F_{\mu\nu}[A]$ and $H_{\mu\nu}[B]$ are not independent but actually related by Hodge dualization:

\begin{align}
^{*}\begin{pmatrix}
F_{\mu\nu} \\
H_{\mu\nu} 
\end{pmatrix}={\cal{S}}\begin{pmatrix}
F_{\mu\nu}\\
H_{\mu\nu} 
\end{pmatrix}, \ \ \ {\cal{S}}=\begin{pmatrix}
0 & 1\\
-1 & 0 
\end{pmatrix}.\label{tsdmax}
\end{align}
Clearly, equation (\ref{tsdmax}) implies the second-order Maxwell equations (\ref{electrom}) by taking the divergence. However, in this first-order formulation $F_{\mu}[A]$ and $H_{\mu\nu}[B]$ appear on an equal footing, related to each other by Hodge dualization. 
\medskip

As a caveat, we notice a redundancy in the twisted self-duality equations (\ref{tsdmax}), for either row can be obtained from the other by Hodge dualization. It is actually possible to identify a non-covariant subset of (\ref{tsdmax}) that is equivalent to the original set of equations and free from redundancies \cite{abh}. This is achieved by selecting the purely spatial components of (\ref{tsdmax}), which produces

\begin{align}
\begin{pmatrix}
{\cal{B}}_{i}[A] \\
{\cal{B}}_{i}[B] 
\end{pmatrix}={\cal{S}}\begin{pmatrix}
{\cal{E}}_{i}[A]\\
{\cal{E}}_{i}[B]
\end{pmatrix},\label{tsdnon}
\end{align}
with ${\cal{E}}_{i}$ and ${\cal{B}}_{i}$ the usual electric and magnetic fields. 

\medskip

Let us now turn the discussion to linearized conformal gravity. Since the dual Weyl tensor $^{*}W_{\mu\nu\r\s}$ has the same algebraic and differential properties as the Weyl tensor, it can be written itself in the same functional form as $W_{\mu\nu\r\s}[h]$ for some different metric $f_{\mu\nu}$:

\begin{align}
&^{*}W_{\mu\nu\r\s}[h] = H_{\mu\nu\rho\sigma}[f],
\end{align}
with

\begin{align}
&H^{\mu}_{\ \nu\r\s}[f] \equiv \tensor{R}{^\mu_\nu_\r_\s}[f]-2(\delta^{\mu}_{\ [\r}S_{\s]\nu}[f]-\delta_{\nu[\r}S_{\s]}^{\ \mu}[f]),
\end{align}
the relation between $h_{\mu\nu}$ and $f_{\mu\nu}$ being non-local. Similarly, it is easy to verify that the Cotton tensor and its Hodge dual have the same properties, and therefore we can write

\begin{equation}
^{*}C_{\mu\nu\r}[h]=D_{\mu\nu\r}[f],
\end{equation}
where $D_{\m\nu\r}[f]$ has the same functional form as the Cotton tensor for the dual metric $f_{\mu\nu}$

\begin{align}
&D_{\mu\nu\r}[f]= -\del_{\r}H^{\r}_{\ \mu\nu\r}[f].
\end{align}
Exactly in parallel as what happens in electromagnetism, Hodge duality exchanges equations of motion and differential identities. In the theory defined by $h_{\mu\nu}$, $\partial_{\mu}\partial_{\nu}W^{\mu\nu\r\s}[h]=0$  is an equation of motion and $\partial_{\mu}\partial_{\nu}^{*}W^{\mu\nu\r\s}[h]$ can be seen as the analogue of the Bianchi identity in Maxwell theory. However, when we consider the dual theory defined by $f_{\mu\nu}$, the latter implies the equation of motion for the dual metric $\partial_{\mu}\partial_{\nu}H^{\mu\nu\r\s}[f]=0$, whereas the former is related to the differential identity $\partial_{\mu}\partial_{\nu}\,^{*}H^{\mu\nu\r\s}[f]=0$.

\medskip

Although the set of equations (\ref{WWW}) is symmetric under the formal exchange of the Weyl tensor and its Hodge dual, implicitly we have prioritized the formulation based on $h_{\mu\nu}$, for $f_{\mu\nu}$ does not appear at all. Following the same logic as discussed above in the case of electromagnetism, it is possible though to find a set of second-order equations equivalent to (\ref{WWW}) where both metrics appear on an equal footing. This is the twisted self-duality equation for linearized conformal gravity:

\begin{equation}
^*\hspace{-.02cm} \begin{pmatrix} W^{\mu\nu\r\s}[h]\\ H^{\mu\nu\r\s}[f]\\ \end{pmatrix} = {\mathcal S} \begin{pmatrix} W^{\mu\nu\r\s}[h]\\ H^{\mu\nu\r\s}[f]\\ \end{pmatrix}, \; \; \; {\mathcal S}  = \begin{pmatrix} 0&1 \\ -1 & 0 \end{pmatrix}.
\label{tsd}
\end{equation}
Equation (\ref{tsd}) is obtained from (\ref{WWW}) by solving for $W_{\mu\nu\r\s}$ and $H_{\mu\nu\r\s}\equiv ^{*}W_{\mu\nu\r\s}$, treated as independent field strengths, and imposing the condition that they are actually related by Hodge dualization. On the other hand, taking the double divergence on both sides, the twisted self-duality equation (\ref{tsd}) reproduces immediately the equations of motion for $W_{\mu\nu\r\s}[h]$ and $H_{\mu\nu\r\s}[f]$ in virtue of the differential identities satisfied by their Hodge duals, exactly as what happens in the Maxwell theory. We notice (\ref{tsd}) also implies the vanishing of the trace of $W^{\mu\nu\r\s}[h]$ and $H^{\mu\nu\r\s}[f]$, owing to the respective cyclic identities. This is in consistency with their definitions as the traceless part of the Riemann tensor for the corresponding metrics $h_{\mu\nu}$ and $f_{\mu\nu}$.

\medskip

\medskip

The set of equations (\ref{tsd}) is redundant, in the sense that either row can be obtained as the Hodge dual of the other one. So we can keep only the subset of equations associated to the first row in (\ref{tsd}):

\begin{align}
&^{*}W^{0i0j} = H^{0i0j}\nn\\
&^{*}W^{0ijk} = H^{0ijk}\nn\\
&^{*}W^{ijkl} = H^{ijkl}\label{set}.
\end{align}
Moreover, we see that the third equation in (\ref{set}) can be obtained from the second one, for

\begin{eqnarray}
^{*}W^{ijkl} = H^{ijkl} \Leftrightarrow \epsilon^{ij0m}W_{0m}^{\ \ kl} = H^{ijkl}\nn\\
\Leftrightarrow W^{0mkl} = -\frac{1}{2}\epsilon_{ij}^{\ \ 0m}H^{ijkl} = -^{*}H^{0mkl},
\end{eqnarray}
the last expression being the Hodge dual of the second equation in (\ref{set}). Thus, the only independent components of the covariant twisted self-duality equations (\ref{tsd}) are

\begin{align}
&^{*}W^{0i0j} = H^{0i0j}\nn\\
&^{*}W^{0ijk} = H^{0ijk}\label{tsd2}.
\end{align}
Defining the electric component $\cE_{ij}$ and magnetic component $\cB_{ij}$ of the Weyl tensor $W_{\mu\nu\r\s}$ as

\begin{align}
\cE_{ij}[h]&\equiv W_{0i0j}[h] \ \ \ \nn\\ 
\cB_{ij}[h]&\equiv -^{*}W_{0i0j}[h]=-\frac{1}{2}\epsilon_{0imn}W_{0j}^{\ \ mn}[h],
\end{align}
and similarly for $H_{\mu\nu\r\s}$

\begin{align}
\cE_{ij}[f]&\equiv H_{0i0j}[f], \ \ \ \nn\\ 
\cB_{ij}[f]&\equiv -^{*}H_{0i0j}[f]=-\frac{1}{2}\epsilon_{0imn}H_{0j}^{\ \ mn}[f],
\end{align}
equation (\ref{tsd2}) can be cast in the form

\begin{equation}
\begin{pmatrix} \cE^{ij}[h]\\ \cE^{ij}[f]\\ \end{pmatrix} = {\mathcal S} \begin{pmatrix} \cB^{ij}[h]\\ \cB^{ij}[f]\\ \end{pmatrix}, \; \; \; {\mathcal S}  = \begin{pmatrix} 0&1 \\ -1 & 0 \end{pmatrix}.
\label{tsd3}
\end{equation}
This equation is non-redundant and contains all the information in the covariant twisted self-duality equation (\ref{tsd}). It will be referred to as the non-covariant twisted self-duality equation. 

\medskip

From their definitions, it is straightforward to see that the electric and magnetic components are both symmetric and traceless. Moreover, their double divergence vanishes:

\begin{eqnarray}
\partial_{i}\partial_{j}\cE_{ij}[h] = \partial_{i}\partial_{j}\cE_{ij}[f] = \partial_{i}\partial_{j}\cB_{ij}[h] = \partial_{i}\partial_{j}\cB_{ij}[f]=0.\nn\\
\end{eqnarray}

\section{Hamiltonian formulation}
\setcounter{equation}{0}

Having established the twisted self-duality structure underlying the equations of motion of linearized conformal gravity, the natural next step is to seek a formulation of the corresponding action principle that manifestly displays duality symmetry. In order to do so, we shall follow the same strategy as in Maxwell theory \cite{deser1} and linearized gravity \cite{HT}: the Hamiltonian formulation is introduced by a 3+1 slicing of space-time, constraints are identified and solved in terms of potentials, and finally a manifest duality symmetric action is written down upon substitution in terms of potentials. This section deals with the Hamiltonian formulation of the theory and the identification of the constraints.

\medskip

The action principle for linearized Weyl gravity is

\begin{align}
S=-\frac{1}{4}\int\;d^{4}xW_{\mu\nu\r\s}W^{\mu\nu\r\s}.
\end{align}
The squared Weyl tensor is decomposed upon a $3+1$ slicing of space-time as follows:

\begin{align}
W_{\mu\nu\r\s}W^{\mu\nu\r\s} &= 4W_{0i0j}W^{0i0j}+4W_{0ijk}W^{0ijk}+W_{ijkl}W^{ijkl}\nn\\ 
&=  8W_{0i0j}W^{0i0j}+4W_{0ijk}W^{0ijk} \label{ld}.
\end{align}
In order to deal with the second-order character of the Lagrangian, we shall follow the Ostrogradski method: to define a dynamical variable depending on first-order derivatives that will be formally treated as independent, and impose afterward its definition through a Lagrange multiplier. A natural choice is the linearized extrinsic curvature:

\begin{align}
K_{ij}=\frac{1}{2}(\dot{h}_{ij}-\partial_{i}h_{0j}-\partial_{j}h_{0i}).
\end{align}
It will be required then to express the Weyl tensor in terms of $K_{ij}$. 

\medskip

First, let us write the Riemann tensor and its contractions in terms of the extrinsic curvature. The components of (\ref{riemann}) are:
\begin{align}
&R_{ijkl}=-\frac{1}{2}[\partial_{i}\partial_{k}h_{jl}+\partial_{j}\partial_{l}h_{ik}-\partial_{i}\partial_{l}h_{jk}-\partial_{j}\partial_{k}h_{il}]\nn\\
&R_{0i0j}=-\dot{K}_{ij}-\frac{1}{2}\partial_{i}\partial_{j}h_{00}\nn\\
&R_{0ijk}=-(\partial_{j}K_{ik}-\partial_{k}K_{ij}).
\end{align}
In turn, the components of the Ricci tensor read

\begin{align}
&R_{00}=R_{0i0}^{\ \ \ i} = -\dot{K}-\frac{1}{2}\Delta h_{00}\nn\\
&R_{0i}=\partial^{k}K_{ik}-\partial_{i}K\nn\\
&R_{ij}=-\dot{K}_{ij}-\frac{1}{2}\partial_{i}\partial_{j}h_{00}+R_{ikj}^{\ \ \ k},
\end{align}
and the scalar curvature is 
\begin{align}
R=-2R_{0i0}^{\ \ \ i}+R_{ij}^{\ \ ij} = 2\dot{K} + \Delta h_{00} + R_{ij}^{\ \ ij}.
\end{align}
From the definition of the Weyl tensor (\ref{weyltensor}) one derives the relations


\begin{align}
W_{0i0j}&=\frac{1}{2}(R_{0i0j} + R_{imj}^{\ \ \ m}) - \frac{1}{6}\delta_{ij}(R_{mn}^{\ \ mn} + R_{0m0}^{\ \ \ m}) \nn\\
&=\frac{1}{2}(-\dot{K}_{ij}-\frac{1}{2}\partial_{i}\partial_{j}h_{00} + R_{imj}^{\ \ \ m})\nn\\
& - \frac{1}{6}\delta_{ij}(-\dot{K}-\frac{1}{2}\Delta h_{00} + R_{mn}^{\ \ mn})\nn
\end{align}
and

\begin{align}
W_{0ijk}&=\partial_{k}K_{ij}-\partial_{j}K_{ik}\nn\\ 
&+ \frac{1}{2}(\delta_{ij}(\partial^{l}K_{lk}-\partial_{k}K)-\delta_{ik}(\partial^{l}K_{jl}-\partial_{j}K)).\nn
\end{align}
                                

The Lagrangian reads

\begin{align}
{\cal{L}}&=-\frac{1}{4}W_{\mu\nu\r\s}W^{\mu\nu\r\s}=-\left[\frac{1}{2}(\dot{K}_{ij}\dot{K}^{ij}+R_{imj}^{\ \ \ m}R^{inj}_{\ \ \ n}\right.\nn\\
&+\dot{K}_{ij}\partial^{i}\partial^{j}h_{00}-2\dot{K}^{ij}R_{imj}^{\ \ \ m}-\partial_{i}\partial_{j}h_{00}R^{imj}_{\ \ \ m})\nn\\
&-\frac{1}{6}(\dot{K}^2+R_{ij}^{\ \ ij}R_{mn}^{\ \ mn}+\dot{K}\Delta h_{00}-2\dot{K}R_{ij}^{\ \ ij}-\Delta h_{00}R_{mn}^{\ \ mn})\nn\\
&\left.+\frac{1}{12}\Delta h_{00}\Delta h_{00}-W_{0ijk}W_{0}^{\ ijk}\right].
\end{align}
The conjugate momentum associated to $K_{ij}$ is defined as usual:

\begin{align}
&P^{ij}=\frac{\partial L}{\partial \dot{K}_{ij}}=-\left[\dot{K}^{ij}+\frac{1}{2}\partial^{i}\partial^{j}h_{00}-R^{imj}_{\ \ \ m}\right.\nn\\
&\left.-\frac{1}{6}\Delta h_{00}\delta^{ij}+\frac{1}{3}R_{mn}^{\ \ mn}\delta^{ij}-\frac{1}{3}\dot{K}\delta^{ij}\right]\label{conj}.
\end{align}
We notice that the trace of $P^{ij}$ vanishes identically:

\begin{align}
P=0\label{c1}.
\end{align}
This shall be treated as a primary constraint. 


\medskip

The Hamiltonian is now introduced through the Legendre transformation

\begin{equation}
{\cal{H}} = P^{ij}\dot{K}_{ij} - {\cal{L}}.
\end{equation}
Upon substitution for the generalized velocities, it can be expressed in terms of $P^{ij}$ and $K_{ij}$: 

\begin{equation}
{\cal{H}} = -\frac{1}{2}P_{ij}P^{ij}-\frac{1}{2}P^{ij}\partial_{i}\partial_{j}h_{00} + R_{imj}^{\ \ \ m}P^{ij} - W_{0ijk}W_{0}^{\ ijk}.
\end{equation}

\medskip

Now we have to take into account the fact that the definition of $K_{ij}$ actually depends on $\dot{h}_{ij}$ by introducing in the action principle the constraint term $\lambda^{ij}(\dot{h}_{ij}-\partial_{j}h_{0i}-\partial_{i}h_{0j}-2K_{ij})$. The factor $\lambda^{ij}$ becomes a Lagrange multiplier enforcing the definition of $K_{ij}$ :

\begin{align}
&S[h_{ij}, p^{ij}, K_{ij}, P^{ij}]=\int\;d^{4}x[P^{ij}\dot{K}_{ij}+\frac{1}{2}Pi_{ij}Pi^{ij}\nn\\
&+\frac{1}{2}P^{ij}\partial_{i}\partial_{j}h_{00}-P^{ij}R_{imj}^{\ \ \ m}\nn\\
&+2\partial_{j}K_{ik}\partial^{j}K^{ik}+2\partial_{k}K\partial_{l}K^{kl}-\partial_{l}K\partial^{l}K-3\partial ^{l}K_{kl}\partial_{m}K^{km}\nn\\
&+\lambda^{ij}(\dot{h}_{ij}-\partial_{j}h_{0i}-\partial_{i}h_{0j}-2K_{ij})]\label{action2}.
\end{align}
Clearly one can identify the Lagrange multiplier $\lambda^{ij}$ with the conjugate momentum associated to $h_{ij}$, $p^{ij}\equiv \lambda^{ij}$. Upon integration by parts, the components $h_{00}$ and $h_{0j}$ act now as Lagrange multipliers imposing the constraints 

\be
\partial_{i}\partial_{j}P^{ij} = 0 \label{ccc1}
\ee
and

\be
\partial_{j}p^{ij}=0.\label{ccc2}
\ee
The latter reads exactly as in linearized gravity \cite{HT}. Ignoring total derivatives coming from the previous integration by parts, the final form of the Hamiltonian action principle is

\begin{align}
&S[h_{ij}, p^{ij}, K_{ij}, P^{ij}, h_{0i}, h_{00}] = \int d^{4}x\left[P_{ij}\dot{K}_{ij} + p^{ij}\dot{h}_{ij}\right.\nn\\ 
&\left.- \cH_{0} + 2\partial_{j}p^{ij}h_{0i} - \frac{1}{2}\partial_{i}\partial_{j}P^{ij}h_{00}\right]
\end{align}
with

\begin{align}
&- \cH_{0} = \frac{1}{2}P^{ij}P_{ij} - P^{ij}R_{imj}^{\ \ \ m} - 2p^{ij}K_{ij}\nn\\ 
&+ 2\partial_{j}K_{ik}\partial^{j}K^{ik} + 2\partial_{k}K\partial^{l}K_{kl} - \partial_{l}K\partial^{l}K - 3\partial^{l}K_{kl}\partial_{m}K^{km}\label{Ham}.
\end{align}
One notices in (\ref{Ham}) the presence of terms linear in the conjugate momenta, which points out the Ostrogradski linear instability of the theory. 
\medskip

There is an additional constraint that comes about by demanding the preservation of the constraint (\ref{c1}) under time evolution. In other words, the Poisson bracket of the constraint (\ref{c1}) with the Hamiltonian should vanish:

\begin{align}
\{P, \int d^3x\,{\cal{H}}\} = 0. 
\end{align}
This results in the constraint

\begin{align}
p = 0 \label{c2}.
\end{align}
The consistency condition applied to this constraint does not produce any further ones.

\medskip

Adding up the traceless constraints (\ref{c1}) and (\ref{c2}), the action principle reads

\begin{align}
&S[h_{ij}, p^{ij}, K_{ij}, P^{ij}, h_{0i}, h_{00}, \lambda_{1}, \lambda_{2}] \nn\\
&= \int d^{4}x\left[P_{ij}\dot{K}_{ij} + p^{ij}\dot{h}_{ij} - {\cal{H}}_{0}+ 2\partial_{j}p^{ij}h_{0i} - \frac{1}{2}\partial_{i}\partial_{j}P^{ij}h_{00}\right.\nn\\
&\left. + \lambda_{1}P + \lambda_{2}p\right]. \label{actionfin}
\end{align}
The gauge transformations of the dynamical variables are

\begin{align}
&\delta h_{ij} = \partial_{i}\xi_{j} + \partial_{j}\xi_{i} + \delta_{ij}\xi\nn\\
&\delta K_{ij} = \frac{1}{2}[-2\partial_{i}\partial_{j}\xi_{0} + \delta_{ij}\dot{\xi}]\nn\\
&\delta p^{ij} = 0\nn\\
&\delta P^{ij} = 0.
\end{align}
These can be obtained directly from the definition of the dynamical variables in terms of the components of the four-dimensional metric $h_{\mu\nu}$. It is straightforward to verify that the constraints (\ref{cc1}), (\ref{cc2}), (\ref{c1}) and (\ref{c2}) are first class, so the previous gauge transformations can also be derived from the Poisson bracket with the constraints. We notice that $p=0$ generates the Weyl rescaling for $h_{ij}$, whereas $\partial_{j}p^{ij}$ is responsible for the same three-dimensional diffeomorphism invariance of linearized Einstein gravity.   



\section{Resolution of the constraints}

We shall now focus on the resolution of the differential constraints
\begin{equation}
\partial_{i}\partial_{j}P^{ij} = 0\label{cc1}
\end{equation}
and

\begin{equation}
\partial_{j}p^{ij} = 0\label{cc2},
\end{equation}
subject to the traceless constraints

\begin{equation}
P = 0\label{t1}
\end{equation}
and

\begin{equation}
p = 0.\label{t2}
\end{equation}
Let us first focus on (\ref{cc1}). Taking into account that $P = 0$, this can be solved in terms of some tensor potential  $\psi_{ij}$ as follows:


\begin{equation}
P^{ij}=\epsilon_{imn}\partial^{m}\psi^{nj} + \epsilon_{jmn}\partial^{m}\psi^{ni}.
\end{equation}
Because of the traceless condition on $P^{ij}$ , the antisymmetric component of  $\psi_{ij}$ is restricted to have the form
\begin{equation}
\psi_{[ij]} = \partial_{i}w_{j} - \partial_{j}w_{i}.
\end{equation}
However, by the redefinition of the symmetric component of the potential  $\psi_{(ij)}\equiv \phi_{ij} + \partial_{i}w_{j} + \partial_{j}w_{i}$ the solution simply takes the form

\begin{equation}
P^{ij} = \epsilon_{imn}\partial^{m}\phi^{nj} + \epsilon_{jmn}\partial^{m}\phi^{ni}\label{res1},
\end{equation}
with $\phi_{ij}$ a symmetric tensor. Note that, since $\delta P^{ij} = 0$, the ambiguities in the definition of the potential $\phi_{ij}$ are restricted to have the form

\begin{equation}
\delta\phi_{ij} = \partial_{i}\partial_{j}\xi + \delta_{ij}\theta.
\end{equation}
This has exactly the same form as $\delta K_{ij}$, which already suggests that $K_{ij}$ and $\phi_{ij}$ can be treated on equal footing, and justifies the renaming $\phi_{ij}\equiv \tilde{K}_{ij}$.

\medskip


In order to solve the constraint (\ref{cc2}), we may use the Poincar\'e lemma and write

\begin{equation}
p^{ij} = \epsilon_{imn}\partial^{m}\omega^{nj} + \epsilon_{jmn}\partial^{m}\omega^{ni}\label{pij}
\end{equation}
for some symmetric potential $\omega_{ij}$, bearing in mind the fact that $p^{ij}$ is symmetric and traceless. However, the additional condition $\partial_{i}p^{ij}=0$ must be fulfilled, which implies that 

\be
\partial_{i}\epsilon_{jmn}\partial^{m}\omega^{ni}=0\label{identity}
\ee
should be identically satisfied. A particular choice of $\omega_{ij}$ that fulfills this condition is

\begin{equation}
\omega_{ij} = R_{ij}[\tilde{h}], \label{choice}
\end{equation}
with $R_{ij}[\tilde{h}]$ having the functional form of the linearized three-dimensional Ricci tensor for some symmetric tensor $\tilde{h}_{ij}$ --to be interpreted in the sequel as a second, dual metric. Indeed, we see that

\begin{equation}
\partial_{i}\epsilon_{jmn}\partial^{m}R^{ni}[\tilde{h}] = \epsilon_{jmn}\partial^{m}(\Delta\partial_{i} \tilde{h}_{ik} - \partial_{l}\Delta \tilde{h}_{kl}) = 0.
\end{equation}
More generally, the condition (\ref{identity}) is identically satisfied for
\be
\omega_{ij} = R_{ij}[\tilde{h}] + \delta_{ij}s_{1} + \partial_{i}\partial_{j}s_{2}. \label{omegaij}
\ee
with $s_{1}$ and $s_{2}$ undetermined scalar functions. The contribution to (\ref{pij}) from the last two terms in (\ref{omegaij}) vanishes, so they can be ignored in practice. 

\medskip

We shall write then
\begin{equation}
p^{ij} = -\frac{1}{2}\left(\epsilon_{imn}\partial^{m}R^{nj}[\tilde{h}] + \epsilon_{jmn}\partial^{m}R^{ni}[\tilde{h}]\right),\label{res2}
\end{equation}
where the global factor has been chosen for future convenience. This expression is invariant under transformations of $\tilde{h}_{ij}$ having the same form as those defining the symmetries of conformal gravity,

\begin{align}
\delta \tilde{h}_{ij} = \partial_{i}\chi_{j} + \partial_{j}\chi_{i} + \delta_{ij}\chi,
\end{align}
as we could have expected.

\medskip

\section{Manifest duality invariance}
\setcounter{equation}{0}

We can now implement equations (\ref{res1}) and (\ref{res2}) in the action principle (\ref{actionfin}). Let us first we compute the quadratic term in $P^{ij}$:

\begin{align}
&\frac{1}{2}P^{ij}P_{ij} = 2\partial_{j} \tilde{K}_{ik}\partial^{j}\tilde{K}^{ik} + 2\partial^{k}\tilde{K}\partial^{l}\tilde{K}_{kl} \nn\\
&- \partial_{l}\tilde{K}\partial^{l}\tilde{K} - 3\partial^{l}\tilde{K}_{kl}\partial_{m}\tilde{K}^{km}.
\end{align}
Remarkably, this has exactly the same form as the quadratic terms in $K_{ij}$ appearing in the Hamiltonian (\ref{Ham}), and suggests an invariance under the transformation

\begin{align}
&K_{ij} \rightarrow \tilde{K}_{ij}, \nn\\
&\tilde{K}_{ij} \rightarrow -K_{ij}. \label{d1}
\end{align}
The kinetic term $P^{ij}\dot{K}_{ij}$ is also invariant under (\ref{d1}) (up to total derivatives):

\begin{align}
&P^{ij}\dot{K}_{ij} = 2\epsilon_{imn}\partial_{m}\tilde{K}_{jn}\dot{K}^{ij} \rightarrow -2\epsilon_{imn}\partial_{m}K_{jn}\dot{\tilde{K}}^{ij}\nn\\
& = 2\epsilon_{imn}\partial_{m}\tilde{K}_{jn}\dot{K}^{ij} + \text{total derivatives}.
\end{align}


\medskip

Substituting now in the term $-2p^{ij}K_{ij}$, we find

\begin{align}
-2p^{ij}K_{ij} &= 2\epsilon_{imn}\partial^{m}R^{nj}[\tilde{h}]K_{ij} = -2\epsilon_{imn}R^{nj}[\tilde{h}]\partial_{m}K_{ij}\nn\\
&+\text{total derivatives}.
\end{align}
We may compare this expression with the term

\begin{equation}
-P^{ij}R_{ij}[h] = -2\epsilon_{imn}\partial_{m}\tilde{K}_{nj}R_{ij}[h]
\end{equation}
and see that these two are rotated into each other by the transformation (\ref{d1}) supplemented by


\begin{align}
&h_{ij}\rightarrow \tilde{h}_{ij}\nn\\
&\tilde{h}_{ij}\rightarrow -h_{ij}.\label{duality}
\end{align}
The kinetic term $\dot{h}_{ij}p^{ij}=-\dot{h}_{ij}\epsilon_{imn}\partial_{m}R_{nj}[\tilde{h}]$ is also invariant under (\ref{duality}) up to total derivatives:

\begin{align}
&-\dot{h}_{ij}\epsilon_{imn}\partial_{m}R_{nj}[\tilde{h}] = \frac{1}{2}\dot{h}_{ij}\epsilon_{imn}\partial_{m}(\Delta \tilde{h}_{nj} - \partial_{j}\partial_{k} \tilde{h}_{kn}) \nn\\
&\rightarrow -\frac{1}{2}\dot{\tilde{h}}_{ij}\epsilon_{imn}\partial_{m}(\Delta h_{nj} - \partial_{j}\partial_{k} h_{kn}) \nn\\
&= \frac{1}{2}\dot{h}_{ij}\epsilon_{imn}\partial_{m}(\Delta \tilde{h}_{nj} - \partial_{j}\partial_{k} \tilde{h}_{kn})+\text{total derivatives}.
\end{align}
So we conclude that, once the constraints are solved, the action principle (\ref{actionfin}) can be cast in the manifestly duality invariant form

\begin{align}
&S[h_{ij},\tilde{h}_{ij},K_{ij},\tilde{K}_{ij},] = \int d^{4}x [2\epsilon_{imn}\partial_{m}\tilde{K}_{jn}\dot{K}_{ij}\nn\\
& - \dot{h}_{ij}\epsilon_{imn}\partial_{m}R_{nj}[\tilde{h}] - \cH],\label{dualityactionfin}
\end{align}
where 

\begin{align}
&-\cH = 2\epsilon_{imn}\partial^{m}\tilde{K}_{nj}R_{ij}[h] - 2\epsilon_{imn}\partial^{m}K_{ij}R_{nj}[\tilde{h}] \nn\\
& 2\partial_{j} K_{ik}\partial^{j}K^{ik} + 2\partial^{k}K\partial^{l}K_{kl} - \partial_{l}K\partial^{l}K - 3\partial^{l}K_{kl}\partial_{m}K^{km}\nn\\
&+ 2\partial_{j}\tilde{K}_{ik}\partial^{j}\tilde{K}^{ik} + 2\partial^{k}\tilde{K}\partial^{l}\tilde{K}_{kl} - \partial_{l}\tilde{K}\partial^{l}\tilde{K} - 3\partial^{l}\tilde{K}_{kl}\partial_{m}\tilde{K}^{km}\nn\\
\end{align}
and we have dropped surface terms. One can verify that the action principle (\ref{dualityactionfin}) is actually invariant under continuous duality rotations of the dual metrics and extrinsic curvatures:

\begin{align}
\begin{pmatrix}
h_{ij} \\
\tilde{h}_{ij} 
\end{pmatrix}&=\begin{pmatrix}\text{cos}\,\theta & \text{sin}\,\theta\\
-\text{sin}\,\theta & \text{cos}\,\theta \end{pmatrix}\begin{pmatrix}
h_{ij}\\
\tilde{h}_{ij} 
\end{pmatrix}\nn\\
\nn\\
\begin{pmatrix}
K_{ij} \\
\tilde{K}_{ij} 
\end{pmatrix}&=\begin{pmatrix}\text{cos}\,\theta & \text{sin}\,\theta\\
-\text{sin}\,\theta & \text{cos}\,\theta \end{pmatrix}\begin{pmatrix}
K_{ij}\\
\tilde{K}_{ij} 
\end{pmatrix}.
\end{align}

\medskip

\section{Conclusions}
\setcounter{equation}{0}


We have shown that electric-magnetic duality is a hidden symmetry of linearized conformal gravity, both at the level of the equations of motion and the action principle. In order to render the symmetry manifest, i.e. to establish a formulation where the ``electric'' and ``magnetic'' degrees of freedom appear on equal footing, it seems necessary to work in a non-manifestly space-time covariant framework. The covariant equations of motion and differential identities obeyed by the Weyl tensor and its Hodge dual can be recovered from a non-covariant subset of the twisted self-duality equation, where the electric and magnetic components of the Weyl tensors for two dual metrics appear on equal footing. The action principle is cast in a manifestly duality-invariant form as well, upon resolution of the constraints in the Hamiltonian formalism. The potentials that solve these constraints are interpreted as a dual three-dimensional metric and a dual extrinsic curvature. Duality acts as simultaneous rotations in the respective spaces spanned by the two metrics and the two extrinsic curvatures. 

\medskip

There are several interesting directions for future work to be discussed. An important question is to determine whether a manifestly duality invariant action principle can be obtained upon linearization around more general backgrounds, in particular (anti) de Sitter space-time. The precise relation between the equations of motion obtained from the duality-symmetric action principle and the non-covariant twisted self-duality equation should be determined. Supersymmetric extensions can also be investigated, along the lines of the work \cite{susy}. Although we have not dealt with topological terms, it may be interesting to study the consequences of their presence: for instance, to investigate if they can cancel out the total derivatives produced by integration by parts in the process of rending the action principle in its manifestly duality-invariant form. Manifest space-time covariance of the action principle might be restored upon introduction of auxiliary fields, although those are expected either to enter in a non-polynomial fashion \cite{PST} or to appear in infinite number \cite{D}. Possible obstructions to manifest duality invariance at higher perturbative orders should also be explored \cite{deser2}. 

\medskip

Electric-magnetic duality in abelian Yang-Mills theory has been discussed at the quantum level in the path-integral formulation \cite{witten}. Here one adds to the abelian action $S[A]$ a term $i\int B\wedge dF$ featuring an additional 1-form field $B$, such that integrating over it produces a delta functional $\delta(dF)$ allowing integration over not necessarily closed 2-forms $F$. If we instead integrate over $F$, the partition function written as an integral over $B$ takes the same form as expressed in terms of the original 1-form field $A$, but interchanging the coupling constant $e^2$ and the $\theta$-parameter. Whether a similar analysis can be performed in linearized conformal gravity seems an avenue worth exploring.


\begin{thebibliography}{99}










 



\bibitem{Ehlers}
J. Ehlers, ``Transformation of static exterior solutions of Einstein's gravitational field equations into different solutions by means of conformal mappings,'' in  ``Les Theories relativistes de la gravitation'', Colloques Internationaux du CNRS 91, 275 (1962).

\bibitem{Geroch}
R.~P.~Geroch, ``A Method for generating solutions of Einstein's equations,'' J.\ Math.\ Phys.\  {\bf 12}, 918 (1971); R.~P.~Geroch,  ``A Method for generating new solutions of Einstein's equations. 2,'' J.\ Math.\ Phys.\  {\bf 13}, 394 (1972).
  
\bibitem{Cremmer}
  E.~Cremmer and B.~Julia,
  ``The SO(8) Supergravity,''
  Nucl.\ Phys.\ B {\bf 159}, 141 (1979).

\bibitem{Julia}
B.~Julia, ``Group disintegrations,'' in ``Superspace and Supergravity'', Hawking, S.W., and Ro\u{c}ek, M., eds., Nuffield Gravity Workshop, Cambridge, England, June 22 - July 12, 1980 (Cambridge University Press, Cambridge, U.K.; New York, U.S.A., 1981) p. 331-350; B.~L.~Julia, ``Dualities in the classical supergravity limits: Dualizations, dualities and a detour via (4k+2)-dimensions,'' in *Strings, branes and dualities. Proceedings, NATO Advanced Study Institute, Cargese, France, May 26-June 14, 1997* Ed. L.~Baulieu, P.~Di Francesco, M.~Douglas, V.~Kazakov, M.~Picco, P.~Windey, NATO Sci.Ser.C 520 (1999)
Dordrecht, Netherlands, p. 121-139.

\bibitem{Nicolai}
T.~Damour, M.~Henneaux and H.~Nicolai, ``E10 and a ``small tension expansion'' of M
Theory'', Phys. Rev. Lett. 89 221601 (2002).

\bibitem{West}
  P.~C.~West,
  ``E(11) and M theory,''
  Class.\ Quant.\ Grav.\  {\bf 18}, 4443 (2001).

\bibitem{Lambert}
N.~D.~Lambert, P.~C.~West, 
``Coset Symmetries in Dimensionally Reduced Bosonic String Theory,''
Nucl.\ Phys. \ B {\bf 615}, 117 (2001).


  
\bibitem{HT}
  M.~Henneaux and C.~Teitelboim,
  ``Duality in linearized gravity,''
  Phys.\ Rev.\  D {\bf 71}, 024018 (2005).

	\bibitem{Julialambda} 
  B.~Julia, J.~Levie and S.~Ray,
  ``Gravitational duality near de Sitter space,''
  JHEP {\bf 0511}, 025 (2005).

\bibitem{me}
S.~H\"ortner, ``Manifest gravitational duality near anti de Sitter space-time,'' Front.in Phys. 7 (2019) 188.

\bibitem{deser1}
S.~Deser and C.~Teitelboim,
``Duality transformations of Abelian and non-Abelian gauge fields,''
Phys.\ Rev.\ D 13 1592  (1976).

\bibitem{hs}
M.~Henneaux, S.~H\"ortner, A.~Leonard, ``Higher Spin Conformal Geometry in Three Dimensions and Prepotentials for Higher Spin Gauge Fields,'' JHEP 1601 (2016) 073.






\bibitem{stelle}
K.~Stelle, Phys.~Rev.~D16, 953 (1977).

\bibitem{adler}
S.~L.~Adler, Rev.~Mod.~Phys.~54, 729 (1982).

\bibitem{maldacena}
J.~Maldacena, arXiv:1105.5632.

\bibitem{susy1}
E.~Fradkin and A.~A.~Tseytlin, Phys.~Rept.~119 (1985) 233-362.

\bibitem{dewitt}
D.~Butter, F.~Ciceri, B.~de Wit and B.~Sahoo, ``Construction of all N$=4$ conformal supergravities,''  	Phys.Rev.Lett. 118, 081602 (2017).

\bibitem{witten1}
N.~Berkovits, E.~Witten, JHEP0408, 009 (2004).

\bibitem{adscft}
H.~Liu and A.~A.~Tseytlin, Nucl.~Phys.~B533, 88 (1998); V.~Balasubramanian,  E.~G.~Gimon,  D.~Minic  and J.~Rahmfeld, Phys.~Rev.~D63,  104009  (2001).


\bibitem{deser}
S.~Deser, R.~I.~Nepomechie, Phys. Lett. A97 (1983) 329-332.


\bibitem{abh}
C.~Bunster, M.~Henneaux, ``The action for twisted self-duality,'' Phys.Rev.D 83 (2011) 125015.

\bibitem{mebh}
C.~Bunster, M.~Henneaux, S.~H\"ortner, ``Gravitational Electric-Magnetic Duality, Gauge Invariance and Twisted Self-Duality'' J. Phys. A: Math. Theor. 46 214016.

\bibitem{susy}
C.~Bunster, M.~Henneaux, ``Supersymmetric electric-magnetic duality as a manifest symmetry of the action for super-Maxwell theory and linearized supergravity,'' Phys.~Rev.~D86 (2012) 065018.

\bibitem{PST}
P.~Pasti, D.~Sorokin and M.~Tonin, Phys.~Rev.~D55, 6292 (1997); P.~Pasti, D.~Sorokin and M.~Tonin, Phys.~Rev.~D52 (1995) 4277.

\bibitem{D}
B. McClain, Y.S. Wu, and F. Yu, Nucl. Phys.B343 (1990) 689; C. Wotzasek, Phys. Rev. Letters 66 (1991) 129
F.P. Devecchi and M. Henneaux, Phys. Rev. D54 (1996) 1606; I. Martin and A. Restuccia, Phys. Lett. B323 (1994) 311; N. Berkovits, Phys. Lett. B388 (1996) 743.

\bibitem{deser2}
S.~Deser and D.~Seminara, ``Free spin 2 duality invariance cannot be extended to GR,''  Phys.~Rev.~D71 081502 (2005). 

\bibitem{witten}
E.~Witten, ``On S-duality in abelian gauge theory,'' Selecta Math.1, 383 (1995).


\end{thebibliography}
\end{document}